\pdfoutput=1

\documentclass[8.5pt,twoside,twocolumn]{article}
\usepackage[super,sort&compress,comma]{natbib} 
\usepackage[version=3]{mhchem}
\usepackage{times,mathptmx}
\usepackage{balance} 
\usepackage{amsmath}
\usepackage{multirow}
\usepackage{xhfill} 
\usepackage{graphicx} 
\usepackage{lastpage}
\usepackage[format=plain,justification=raggedright,singlelinecheck=false,font=small,labelfont=bf,labelsep=space]{caption} 
\usepackage{fancyhdr}
\newcommand{\kpoint}{$\textit{\textbf{k}}$-point}

\begin{document}

\thispagestyle{plain}
\fancypagestyle{plain}{
\renewcommand{\headrulewidth}{1pt}}
\renewcommand{\thefootnote}{\fnsymbol{footnote}}
\renewcommand\footnoterule{\vspace*{1pt}%
\hrule width 3.4in height 0.4pt \vspace*{5pt}} 
\setcounter{secnumdepth}{5}

\makeatletter 
\def\subsubsection{\@startsection{subsubsection}{3}{10pt}{-1.25ex plus -1ex minus -.1ex}{0ex plus 0ex}{\normalsize\bf}} 
\def\paragraph{\@startsection{paragraph}{4}{10pt}{-1.25ex plus -1ex minus -.1ex}{0ex plus 0ex}{\normalsize\textit}} 
\renewcommand\@biblabel[1]{#1}            
\renewcommand\@makefntext[1]%
{\noindent\makebox[0pt][r]{\@thefnmark\,}#1}
\makeatother 
\renewcommand{\figurename}{\small{Fig.}~}

\setlength{\arrayrulewidth}{1pt}
\setlength{\columnsep}{6.5mm}
\setlength\bibsep{1pt}

\twocolumn[
  \begin{@twocolumnfalse}
\noindent\LARGE{\textbf{Ab initio thermodynamic model of \ce{Cu2ZnSnS4}}$^{\dag}$}
\vspace{0.6cm}

\noindent\large{\textbf{Adam J. Jackson \textit{$^{a}$} and
Aron Walsh$^{\ast}$\textit{$^{b}$}}}\vspace{0.5cm}



\noindent \normalsize{
Thin-film solar cells based on the semiconductor \ce{Cu2ZnSnS4} (CZTS) are a promising candidate for Terawatt-scale renewable energy generation.
While CZTS is composed of earth abundant and non-toxic elements, arranged in the kesterite crystal structure, there is a synthetic challenge to produce high-quality stoichiometric materials over large areas.
We calculate the thermodynamic potentials of CZTS and its elemental and binary components based on energetic and vibrational data computed using density functional theory.
These chemical potentials are combined to produce a thermodynamic model for the stability of CZTS under arbitrary temperatures and pressures, which provide insights into the materials chemistry.
CZTS was shown to be thermodynamically stable with respect to its component elements and their major binary phases binaries under modest partial pressure of sulfur and temperatures below 1100K.
Under near-vacuum conditions with sulfur partial pressures below 1 Pa decomposition into binaries including solid SnS becomes favourable, with a strongly temperature-dependent stability window.
}
\vspace{0.5cm}
 \end{@twocolumnfalse}
  ]

\section{Introduction}
\footnotetext{\dag~Electronic Supplementary Information (ESI) available: [Material properties and thermodynamic data implemented as Python modules. Source code for free energy surface plots.]. 
}


\footnotetext{\textit{$^{a}$~Doctoral Training Centre in Sustainable Chemical Technologies, University of Bath, Claverton Down, Bath, UK. 
}}
\footnotetext{\textit{$^{b}$~Centre for Sustainable Chemical Technologies and Department of Chemistry, University of Bath, Claverton Down, Bath, UK. 
E-mail: a.walsh@bath.ac.uk}}



Inorganic thin-film solar cells consist of several materials (a combination of metallic, semiconducting and insulating compounds) arranged in a particular order to exploit the photovoltaic effect. 
The deposition and optimisation of each layer requires a specific set of conditions and sometimes chemical treatments to `activate' their performance. 
The role of compositional and structural variations in limiting performance for a range of solar cell technologies is known.\cite{Luque2003} 
The development of processing and annealing conditions has been largely empirical in the past; however, the importance of chemical thermodynamics in this area is beginning to be recognised.\cite{Scragg2012}

\begin{figure}[ht!]
\begin{center}
\includegraphics[width=0.9\columnwidth]{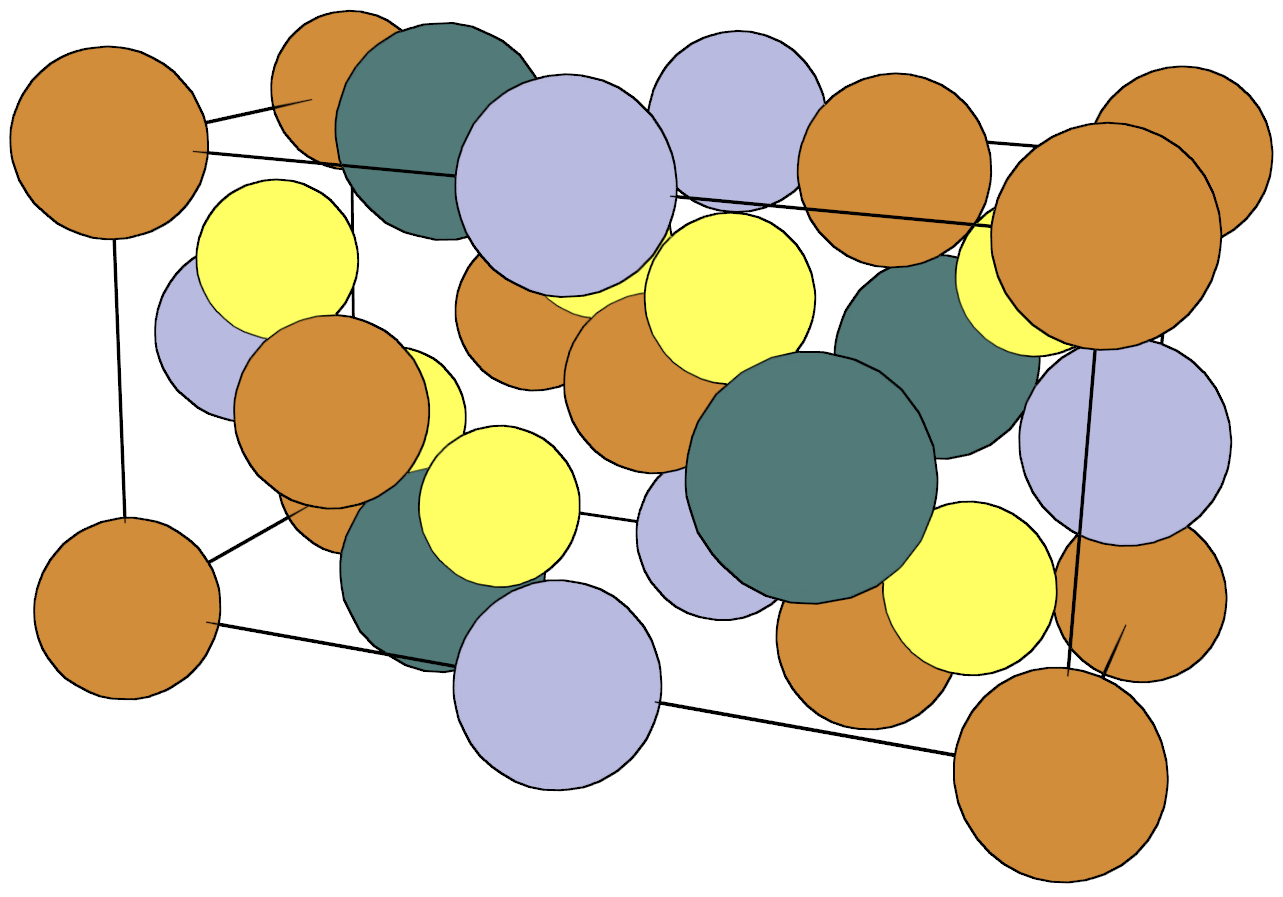} 
\end{center}
\caption{\label{fig-structure} Tetragonal unit cell of kesterite structured CZTS. Alternate layers of Cu (orange) with Zn (blue) and Sn (green) alternate with layers of interstitial sulfur atoms (yellow) in an extended zinc blende-like arrangement of tetrahedral coordination environments.}
\end{figure}

\ce{Cu2ZnSnS4} (CZTS), a quaternary chalcogenide semiconductor, was introduced as a photovoltaic material in 1988 by Ito and Nakazawa.\cite{Ito1988}
Several crystal structures are known, but while early work assumed the stannite structure (space group $I\bar{4}2m$), the lowest-energy structure is now known to be the kesterite structure (space group $I\bar{4}$, Fig \ref{fig-structure}). 
Computational work has shown this to be a few meV per formula unit lower in ground-state energy than the stannite and CuAu-derived structures.\cite{chen-041903}
All low-energy crystal structures are related to the face-centred cubic zincblende lattice, with Cu, Zn and Sn distributed over one sublattice and S filling a second sublattice.

In recent years CZTS and Se-containing variations have come under particular attention as a candidate for large-area thin-film cells, with a current record light-to-electricity conversion efficiency of 12.6 \% in a Se-dominated cell.\cite{wang-201301465}
The record efficiency for Se-free CZTS is 8.4 \%.\cite{Shin2011}
Its distinct advantages over competing technologies are the combination of a direct optical bandgap around the ``optimum'' 1.5 eV and its abundant, inexpensive elemental components.\cite{Ito1988,wang-201301465,Scragg2008,katagiri-2455,chen-041903}
The reserves and production rates of these materials suggest that CZTS is a strong contender for global-scale generation compared to peers including Cu(In,Ga)(S,Se)$_2$, \ce{CdSe} and \ce{CdTe}.\cite{Wadia2009}
The long-term requirements for such generation are expansive, forming part of ``country-sized renewable facilities''.\cite{Mackay2013}
Large-scale production requires a pragmatic process, preferably one which is adaptable to ``roll-to-roll'' processing.
Rapid reactions and modest pressures are therefore of particular interest, as is the avoidance of exotic and dangerous substances.

Laboratory studies have already demonstrated the complexity of the phase diagram, with secondary phases or partial disproportionation commonly observed, 
and off-stoichiometry compositions employed in order to manipulate this.\cite{siebentritt-181905,chen-254204,walsh-400,Mousel2013}
These are commonly expressed in terms of chemical potential, giving insight into the transitions between phases but not the corresponding physical conditions.
In particular, the materials chemistry of the system is known to be sensitive to the partial pressure of the chalcogen atmosphere.
This pressure has been manipulated by supplying S/Se solids, \ce{H2S} gas and/or SnS(e) solids.\cite{Ahmed2012,Weber2010,Guo2010,Katagiri1997,Ericson2013}
A recent paper provided Sn and Se powders with the intent that they would form reactive gas-phase Sn-Se compounds.\cite{Fairbrother2014}
It should be noted that SnS is itself a semiconductor that has been attracting interest for application as an absorber layer in thin-film solar cells.\cite{burton-132111,sinsermsuksaku-053901} 

Weber \textit{et al.} studied the relationship between the composition of CZTS and temperature, finding a significant shift in composition and loss of Sn at temperatures above around 500~$^\circ$C under $10^{-2}$ Pa of S, with some SnS evaporation at temperatures as low as 350~$^\circ$C.\cite{Weber2010}
As well as two reactions involving a ternary phase, they propose the quaternary decomposition to binaries and sulfur vapours:
\begin{equation}
\ce{Cu2ZnSnS4}\text{(s)} \rightarrow \ce{Cu2S}\text{(s)} + \ce{ZnS}\text{(s)}
     + \ce{SnS}\text{(g)} + \ce{S}\text{(g)}
\end{equation}
where the SnS is lost to the vapour phase.
Scragg \textit{et al.} studied the SnS - S interaction experimentally and concluded with the aid of kinetic modelling that this is a two-step reaction in which solid SnS is formed in an equilibrium reaction, liberating sulfur vapours before evaporating to SnS vapour:\cite{Scragg2011a}
\begin{align}
\ce{Cu2ZnSnS4} &\rightleftharpoons \ce{Cu2S} + \ce{ZnS} + \ce{SnS}\text{(s)} + \tfrac{1}{2}\ce{S2}\text{(g)} \label{eqn:scragg-SnS}
\intertext{and}
\ce{SnS}\text{(s)} &\rightleftharpoons \ce{SnS}\text{(g)}.\label{eqn:SnS-vapour}
\end{align}

They found the partial pressure of \ce{S2} to be critical in the region $10^{-4}$~mbar ($10^{-2}$~Pa), predicting a stability envelope given sufficient SnS vapour to prevent irreversible evaporation.

A typical chalcogenide photovoltaic device consists of a metallic back electrical contact, an active p-type absorber layer, an n-type buffer layer and a transparent oxide electrical contact. 
For kesterite-based devices this is usually Mo\textbar{}CZTS\textbar{}CdS\textbar{}\ce{In2O3}:Sn. 
Evidence of chemical reactivity and compositional gradients is frequently found at each of the interfaces.\cite{Scragg2012} 
In particular, significant amounts of \ce{MoS2} is formed at the Mo|CZTS interface during the annealing step.\cite{Scragg2013}
It is known that such reactions change the current-voltage characteristics of the solar cell, but the specific processes occurring and how to control them are not understood.
There are significant opportunities for materials chemists to characterise the structure and properties of photovoltaic systems such as these. 

While reference thermochemical data is available for well-studied semiconductors such as ZnS and SnS$_2$, for CZTS and related compounds such information is currently unknown.\cite{crc,Chase1998}
For CZTS it is difficult to define a reference experimental sample;
a recent study employing scanning transmission electron microscopy suggested substantial cation inhomogeneity in CZTS with features of the order 1nm, while atom probe tomography of CZTSe has revealed a co-existing ZnSe network with features of the order 10nm.\cite{Mendis2014,Schwarz2013}
Significant cation disorder is expected in kesterites following high-temperature annealing due to the gain in configurational entropy, and controlled by largely-undocumented cooling rates.\cite{Scragg2014}

In this study, we combine quantum and statistical mechanics to compute a range of thermodynamic potentials for pure kesterite CZTS and its elemental and binary components. 
This consistent set of data, with energies and vibrations computed using density functional theory (DFT), is used to predict the  stability window for the material with respect to different processing conditions. 
The database is freely available and the model will be extended to include other materials systems and processing scenarios.$^\ddag$
\footnotetext{\ddag~The project is hosted at {\tt http://github.com/WMD-Bath/CZTS-model}  and a current snapshot is included in the ESI.}
It is not practical to examine the effect of long-range disorder with \emph{ab initio} calculations, and if these are suspected to provide a significant thermodynamic driving force in CZTS formation then alternative approaches will be needed.
The stannite phase was briefly examined and the data is included in the ESI$^\dag$, but as the chemical potential is consistently within a few meV of the kesterite phase it does not meaningfully affect any equilibrium results.
A configurational entropy term from inter-mixing is possible, but would require a greater understanding of any phase-coexistence.

%
%

\section{Methodology}
\subsection{Thermodynamic framework}
Classical thermodynamics is used here to predict heats of formation and relative phase stabilities.
A consistent approach is used to calculate the key thermodynamic potentials: internal energy $U$ and enthalpy $H$.
By calculating $U$ as a function of temperature and pressure, the heat capacity $C_v$ and entropy $S$ are also derived, and hence the Helmholtz free energy $A$ and Gibbs free energy $G$.
$G$ may be seen as the `key' to phase stability, as this potential is minimised at equilibrium.

Overall enthalpy changes $\Delta H$ are calculated from the molar enthalpies of components $\hat{H}_i$ following Hess's law:
\begin{equation}
\Delta H = \sum\limits_i \Delta n_i \hat{H}_i
\end{equation}
where $\Delta n_i$ is the stoichiometry change associated with component $i$.
Likewise, Gibbs free energy changes are a sum of species chemical potentials $\mu_i$:
\begin{equation}
\Delta G = \sum\limits_i \Delta n_i \mu{}_i .
\end{equation}
These species-wise energies require a consistent reference point.
Here the reference is a state in which all electrons are non-interacting (i.e., infinitely separated). 
We make the common assumption that chemistry is governed entirely by the electrostatic interactions and kinetic energy of electrons and nuclei, within the Born-Oppenheimer approximation.
Gravity and nuclear forces are neglected.

The chemical potentials can be separated into a ground-state contribution and a vibrational contribution; in this work:
\begin{equation}
U^0 = H^0 = A^0 = G^0 = E^\text{DFT} + E^\text{ZP},
\end{equation}
where a superscript `0' indicates conditions of absolute zero temperature and pressure.
$E^\text{DFT}$ is the (athermal) ground state energy from density functional theory calculations, and $E^\text{ZP}$ is the energy from zero-point vibrations.
Electronic excitations are neglected.

Expressions for these thermochemical potentials as functions of temperature and pressure, 
while aligning suitable reference energies,
 are at the heart of \emph{ab initio} thermodynamics.\cite{Stoffel2010,Reuter2005,Reuter2001,Soon2007}
The forms used here were derived and applied in previous work by the authors\cite{Jackson2013d}: for ideal gases
\begin{align}
\hat{H}_i(T,p_i) &=
	E^\text{DFT} + E^\text{ZP} + \left[ H_i^\theta - H_i^0 \right]
	+ \int^T_{T^\theta} C_p\text{d}T \\
	\begin{split}
\mu_i (T,p_i) &=
	E^\text{DFT} + E^\text{ZP} + \left[ H_i^\theta - H_i^0 \right]
	+ \int^T_{T^\theta}C_p\text{d}T \\
	& \quad - RT\ln \left[p_i/p_i^\theta\right]
	- TS(T,p_i^\theta) ;
	\end{split}
	\intertext{while for incompressible solids} \\
	\begin{split}
\hat{H}_i (T,p_i) &=
	E^\text{DFT} + E^\text{ZP} + \left[ H_i^\theta - H_i^0 \right]\\
	& \quad + \int^T_{T^\theta} C_v\text{d}T	+ PV
	\end{split} \\
	\begin{split}
\mu_i (T,p_i) &= 
	E^\text{DFT} + E^\text{ZP} + \left[ H_i^\theta - H_i^0 \right] \\
	& \quad + \int^T_{T^\theta}C_v\text{d}T
	 + PV - TS (T,p_i^\theta).
	\end{split}
\end{align}
$\theta$ is used to denote an intermediate reference state;
thermodynamic properties are often provided relative to a standard temperature and pressure.
This enables the use of standard heat capacities $C_p$ and $C_v$, or tabulated enthalpies relative to some arbitrary state.
(Note that $C_v$ and $C_p$ are used interchangeably for incompressible solids.)

We emphasise that we cannot comment on the rates of reactions from our thermodynamic treatment.
It is possible, for example, that kinetic barriers for a particular phase separation would be prohibitive for it to complete in the timescale of typical processing or annealing conditions. 

\subsection{Computational details}
Ground-state total energies and forces were computed in DFT calculations with the FHI-aims quantum chemistry code.\cite{Blum2009,Havu2009}
These were used to optimise the initial crystal and molecular structures, without symmetry constraints, before modelling vibrational properties.
The PBEsol exchange-correlation (XC) functional was employed; this functional employs the generalised gradient approximation (GGA) and is optimised for solid-state calculations.\cite{Perdew2008} 
Evenly-spaced \kpoint{} grids were used, with scaled sampling following the procedure of Moreno and Soler.\cite{Moreno1992}
In general a 10~\AA{} reciprocal-space cutoff is sufficient for semiconductors and insulators, while a higher cutoff is helpful for achieving convergence in total energies for metals.
The recommended `tight' set of numerically-tabulated atom-centred basis functions was used throughout, except for the cases of Zn metal in which an extended set of 13 basis functions was employed, and Sn metal in which a full `tier 2' basis set was used (17 functions per atom).

Vibrational calculations were performed with Phonopy$^\ddag{}$,
 a code which implements the Parlinski-Li-Kawazoe ``direct method'' for computing solid-state phonons within the harmonic approximation.\cite{Parlinski1997,Togo2008}
\footnotetext{\ddag{}~Phonopy is an open source code developed by Atsushi Togo from the earlier package FROPHO. It is available from {\tt http://phonopy.sourceforge.net}}
From a primitive crystal structure, a mixture of finite displacements and analytical gradients is used to construct a dynamical matrix of force constants.
Forces are obtained from DFT calculations on large periodic cells; generally it is necessary to form supercells in order to avoid self-interaction of displaced atoms.
Supercell sizes for this study are listed in Table~\ref{table:phonon_supercells}.

In this work the electronic structure was iterated until the analytical forces were converged to within $10^{-5}$~eV/\AA{}.
Finite displacements of 0.01~\AA{} were used, and symmetry employed to reduce the number of calculations where possible.
From the dynamical matrix, a set of frequencies is calculated, forming a vibrational model of the system within the harmonic approximation and defining $E^\text{ZP}$.
By applying Bose-Einstein statistics, a relationship is formed between temperature and vibrational energy, yielding $U(T)$, and hence $C_v(T)$ and $S(T)$.
Note that the effect of pressure is not taken into account with this method;
a logical extension for exploring the anharmonic effects of lattice expansion would be the quasi-harmonic method or thermodynamic integration from molecular dynamics simulations with an appropriate ensemble.

\subsection{Crystal structures}
Ternary phases and metal alloys were disregarded at this initial stage. 
The structure for CZTS was drawn from previous work, and optimised for the basis set and XC functional used in this study.\cite{Chen2010} 

The initial structure of Cu was obtained from the Inorganic Crystal Structure Database (ICSD) in the form of a simple face-centred cubic cell (collection code 64699).\cite{Fletcher1996}
For Zn an initial structure was drawn from the ICSD (collection code 64990) consisting of a 2-atom hexagonal-close-packed unit cell.
Standard local geometry-optimisation algorithms (also tested in the plane-wave code VASP) struggled to find the energy-minimising geometry as Zn is soft in the $c$-axis.
A series of fixed $c$-values were tested over increments of 0.025 \AA{}, while relaxing the $a$ and $b$ parameters to find optimal $a$ and $c$ values; $b$, $\alpha$, $\beta$, $\gamma$ and the atomic positions were then fixed to the $P6_3/mmc$ space group.
\label{sec:Zn-opt}
For $\beta$-Sn the lattice constant was drawn from the ICSD (collection code 40039) and a 2-atom face-centred-cubic cell constructed.
The $\alpha$-S structure was based on a previous DFT study and is a large triclinic cell containing 32 S atoms; this in turn is a symmetry reduction from an orthorhombic conventional cell.\cite{Burton2012a}

The structures of \ce{Cu2S} phases have been previously studied using X-ray crystallography and density functional theory; 
while the high-temperature and cubic structures contain partially-occupied sites, low-\ce{Cu2S} has a well-defined, albeit large, 144-atom monoclinic unit cell.
\cite{Xu2012a}
\ce{SnS2} is the binary phase corresponding to the formal oxidation state in CZTS (i.e. Sn(IV)).
A 3-atom hexagonal unit cell was obtained from the ICSD (collection code 100612).
An 8-atom orthorhombic structure for the stable $Pnma$ phase of SnS was drawn from recent work on the phase stability of this material.\cite{Burton2012a}
\ce{ZnS} is encountered in (and gives its name to) both the zincblende and wurtzite crystal structures.
The starting point was 2-atom zincblende primitive cell with a lattice parameter $a = 5.4053$ \AA{} from reference data.\cite{Madelung2004}
A corresponding set of calculations were performed for the wurtzite phase of ZnS; the results are not considered here as the zinc blende ground state is more stable and preferred for modelling, but the results are available as part of the ESI.$^\dag$
The initial and optimised structural parameters for all other phases are included in Table~\ref{table:stage1_pbesol}.
\begin{table}
\caption{Supercells used for phonon calculations. Vibrational frequencies calculated with 0.01 \AA{} displacements and PBEsol analytical gradients. Cutoff length for evenly-spaced \kpoint{} grids as defined in Ref.~\citenum{Moreno1992}.
\label{table:phonon_supercells}}
\begin{minipage}{\textwidth}
\begin{tabular}{cccc}
\hline
\multirow{2}{*}{Species} & Supercell & Supercell & \kpoint{} \\
& expansion & volume (\AA{}$^3$) & cutoff (\AA{})\\ \hline
Cu & $[3 3 3]$ & 1225.5 & 15 \\
Zn & $[4 4 3]$ & 1356.4 & 25 \\
$\beta$-Sn & $[3 3 3]$ & 1879.5 & 25 \\
$\alpha$-S & $[2 2 2]$ & 6663.3 & 10 \\
\ce{Cu2S} & $[1 1 1]$ & 2055.9 & 10\\
ZnS & $[3 3 3]$ & 1038.3 & 10\\
\ce{SnS} & $[3 3 3]$ & 1679.4 & 15 \\
\ce{SnS2} & $[3 3 2]$ & 1252.0 & 10\\
\ce{Cu2ZnSnS4} & $[2 2 2]$ & 2486.3 & \hspace{4pt}10$^a$\\
\hline
\end{tabular}
\footnotetext{$a$~10 \AA{} cutoff exceeded in $c$ axis: [2 2 2] grid.}
\end{minipage}
\end{table}

\begin{table*}
\newcommand{\matspace}{\vspace{2mm}}
\newcommand{\ditto}[1][.4pt]{\xrfill{#1}~''~\xrfill{#1}}
\begin{center}
\caption{Lattice parameters for CZTS, elemental and binary precursors, before and after unit cell optimisation with the PBEsol functional. 
No symmetry constraints were enforced except for the case of Zn metal (discussed in Section~\ref{sec:Zn-opt}). 
Except where other references are given, initial structures were drawn from the Inorganic Crystal Structure Database and collection codes are given in their respective discussions above.\cite{Fletcher1996} $a$, $b$ and $c$ are the lattice vector lengths in \AA; $\alpha$, $\beta$ and $\gamma$ are the angles between the vectors in degrees.
 Calculated enthalpies of formation, $\Delta H_f^\theta$, are given for compounds at standard conditions of 298.15K and 1 bar pressure.
\label{table:stage1_pbesol}}
\begin{tabular}{cccccccccccccccc}
\hline
\multirow{2}{*}{Material} & \multirow{2}{*}{Structure} & 
	\multirow{2}{*}{Space group} & \multirow{2}{*}{$a$} & \multirow{2}{*}{$b$} & 
	\multirow{2}{*}{$c$} & \multirow{2}{*}{$\alpha$} &
	\multirow{2}{*}{$\beta$} & \multirow{2}{*}{$\gamma$} & $\Delta H_f^\theta$ \\
	 & & & & & & & & & (kJ mol$^{-1}$) \\
\hline 
	\ce{Cu2ZnSnS4} & Initial\cite{Chen2010} &  \multirow{2}{*}{$I$-4}
	&5.434& 5.434&10.856& 90.00& 90.00& 90.00 & \multirow{2}{*}{-369.13}\\
	(kesterite) & Optimised && 5.383& 5.383&10.727& 89.98& 89.99& 89.99 &
	 \matspace \\
	\multirow{2}{*}{Cu} & Initial & \multirow{2}{*}{$Fm$-$3m$}
	& 3.615& 3.615& 3.615& 90.00& 90.00& 90.00 &\\
	& Optimised &&  3.567& 3.567& 3.567& 90.00& 90.00& 90.00 	&
	\matspace \\	
	\multirow{2}{*}{Zn} & Initial & \multirow{2}{*}{$P6_3/mmc$} 
	& 2.665& 2.665& 4.947& 90.00& 90.00&120.00 & \\
	& Optimised &&  2.614& 2.614& 4.775& 90.00& 90.00&120.00 & \matspace \\
	\multirow{2}{*}{$\beta$-Sn} & Initial & \multirow{2}{*}{$I4_1/amd$} 
	&  4.589& 4.589& 4.589& 60.00& 60.00& 60.00 & \\
	& Optimised & & 4.614 & 4.614 & 4.614 & 60.10 & 60.10 & 60.10 & \matspace \\

%
	\multirow{2}{*}{$\alpha$-S} & Initial\cite{Burton2012a} & 
	\multirow{2}{*}{$Fddd$}
	& 14.349&14.237& 7.885& 74.74& 73.21& 32.01 & \\
	& Optimised && 13.788&13.283& 8.335& 75.15& 68.51& 36.02 & \matspace \\
	\multirow{2}{*}{\ce{Cu2S}} & Initial\cite{Xu2012a} & \multirow{2}{*}{P$2_1$/c}
	& 14.424& 11.865& 13.003&  90.00& 116.77&  90.00 & \multirow{2}{*}{-46.24}\\
	& Optimised && 14.870& 11.744& 13.095 & 90.00& 115.97&  90.00 & \matspace \\
	\multirow{2}{*}{\ce{SnS}} & Initial\cite{Burton2012a} & \multirow{2}{*}{$Pnma$}
	& 11.106& 3.989& 4.238& 90.00& 90.11& 90.08 & \multirow{2}{*}{-97.70}\\
	& Optimised && 11.083& 3.982& 4.229& 90.00& 90.00& 90.00  \matspace \\
	\multirow{2}{*}{\ce{SnS2}} & Initial & \multirow{2}{*}{$P$-$3m1$}
	& 3.605& 3.605 & 5.460 & 90.00 & 90.00& 120.00 & \multirow{2}{*}{-120.97}\\
	& Optimised && 3.654 & 3.654 & 6.016 & 89.95 & 90.04 &120.00 &  \matspace \\
	\multirow{2}{*}{ZnS} & Initial & \multirow{2}{*}{$F$-$43m$}
	&  3.822 & 3.822 & 3.822 & 60.00 & 60.00 & 60.00 & \multirow{2}{*}{-156.74}\\
	& Optimised &&  3.789 & 3.789 & 3.789 &59.98 & 59.98 & 59.98 &  \matspace \\
	\hline
\end{tabular}
\end{center}
\end{table*}

\subsection{Sulfur vapours}
The thermochemistry of sulfur has been studied relatively lightly given its abundance and the scale of application.
While sulfur vapours are known to consist of a series of rings from \ce{S8} down to the dimer \ce{S2}, data for the intermediate rings and equilibrium mixtures is relatively scarce.
Standard thermochemical tables prefer to treat the gas phase as an ideal diatomic gas, as do other treatments of the CZTS equilibrium.\cite{Chase1998,Scragg2011a,Scragg2012}
However, the vapour phase of sulfur is thought to contain a mixture of the cyclic allomorphs, and even above 1200 K may only be around 70\% \ce{S2}.\cite{West1950,Berkowitz1963,Rau1973a}

In the condensed phase, the stable structure $\alpha$-S is a molecular solid formed of packed \ce{S8} rings.
In the absence of a comprehensive model, calculations here consider $\alpha$-S solid and the \ce{S2} and \ce{S8} vapours; the true behaviour of the mixture is expected to be somewhere between the effect of these pure species. DFT-optimised structures were used for the ground state energies of the \ce{S2} dimer and \ce{S8} ring.
Temperature-dependent data was drawn from the NIST-JANAF thermochemical data tables; these are based on spectroscopy and assume ideal behaviour.\cite{Chase1998}

\section{Results}

We begin by reporting the vibrational properties for each solid compound of interest.
The resulting thermodynamic potentials are then combined to assess the formation of CZTS with respect to its constituent elements and isovalent binary sulfides. 
Finally the decomposition liberating SnS and sulfur vapour is considered.

\subsection{Lattice dynamics}
Optimised ground-state lattice parameters are given in Table~\ref{table:stage1_pbesol}.
The only significant shifts in structure are for $\alpha$-S (a large, soft, molecular crystal) and to a lesser extent the $c$ parameters of Zn metal and \ce{SnS2}, which are weakly bound.
Following the procedure outlined above, phonon densities of states and band structures were computed for CZTS, Cu, Sn, Zn, $\alpha$-S,  \ce{Cu2S}, ZnS, and \ce{SnS2}. 
While there have been isolated reports of individual phases, this is the first consistent collection of phonon data for these materials.

\begin{figure}
\begin{center}
\includegraphics{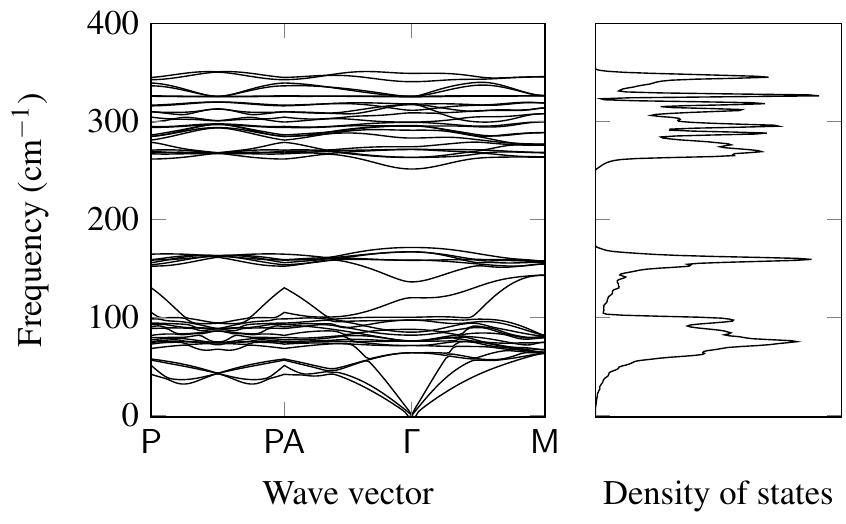}
\caption{Phonon band stucture and density of states for kesterite CZTS (tetragonal unit cell) with 16 atoms and 48 vibrational modes. All other dispersion plots are included in ESI.$^\dag$
\label{fig:CZTS_pdos}}
\end{center}
\end{figure}

The phonon dispersion for CZTS is shown in Fig.~\ref{fig:CZTS_pdos}, while the remaining curves are available in the ESI.$^\dag$
The general behaviour is similar to previous work by Khare \textit{et al.}, but lacks LO-TO splitting at the $\Gamma$ point, which was not considered in this case.\cite{Khare2012} 
The 16-atom unit cell of kesterite results in 48 (3\textit{N}) modes. 
There are two blocks of bands, which spread from 0 to 171~cm$^{-1}$ and from 251 to 350 cm$^{-1}$. 
The calculated phonon density of states (DOS) (Fig.~\ref{fig:CZTS_pdos}) shows activity between around 270 and 250~cm$^{-1}$; from experimental studies the Raman spectrum is known to contain two distinct peaks at about 286 and 337~cm$^{-1}$ (A$_1$ modes).\cite{siebentritt-181905}

The associated total energies and thermal properties have been packaged into a Python code for ease of use and manipulation.
These are also available as supplementary information, and as part of a continuing project online.$^\dag{}$

\subsection{Standard thermodynamic properties of CZTS}

\subsubsection{Heat of formation. } The formation enthalpy is defined with respect to the component elements in their (solid) standard states,
\begin{equation}\label{eqn:dhf}
\ce{2Cu} + \ce{Zn} + \ce{Sn} + \ce{4S} \rightleftharpoons \ce{Cu2ZnSnS4}
\end{equation}
The standard formation enthalpy of kesterite CZTS at 298.15~K and under 1~bar of pressure is calculated to be $-3.83$~eV per formula unit ($-369.1$~kJ~mol$^{-1}$). 
At standard conditions the effect of temperature and pressure is small ($< 1$~meV) as there is no gas component to the reaction.
However, stability is determined by the Gibbs free energy of formation, $\Delta G_f$.

\begin{figure}
\includegraphics[width=\columnwidth]{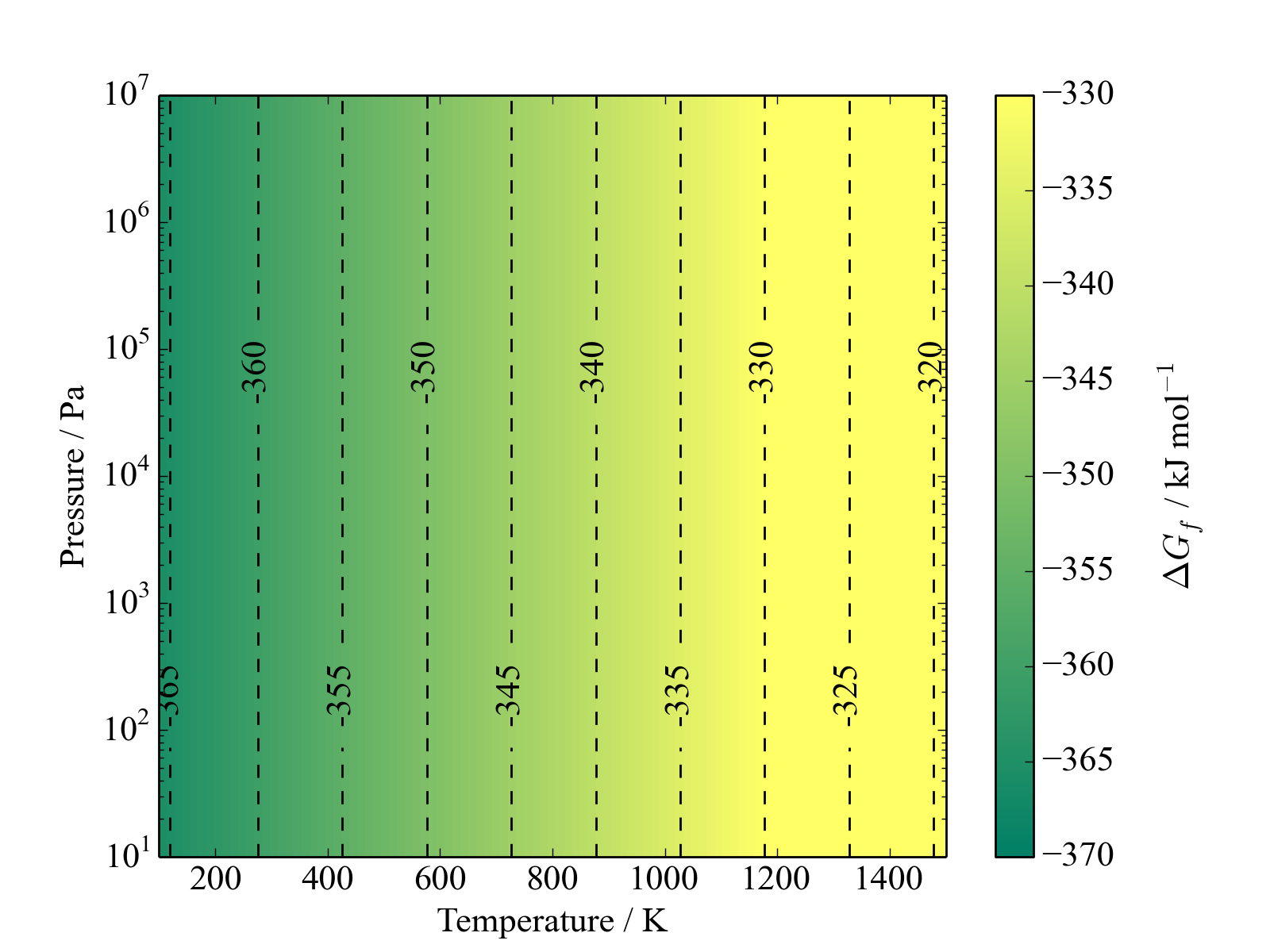} 
\caption{Gibbs free energy of formation of kesterite CZTS from elements in their standard states (Eqn. \ref{eqn:dhf}). Contours represent energies in kJ~mol$^{-1}$. S is in the solid $\alpha$-sulfur phase. The y-axis represents pressure from an inert gas or mechanical force.
\label{fig:DG_CZTS_alpha}
}
\end{figure}

The free energy of formation is plotted against temperature and pressure in Fig.~\ref{fig:DG_CZTS_alpha}.
The effect of pressure is negligible due to assumed solid incompressibility and absence of a gas phase.
\begin{figure}
\begin{center}
\includegraphics[width=\columnwidth]{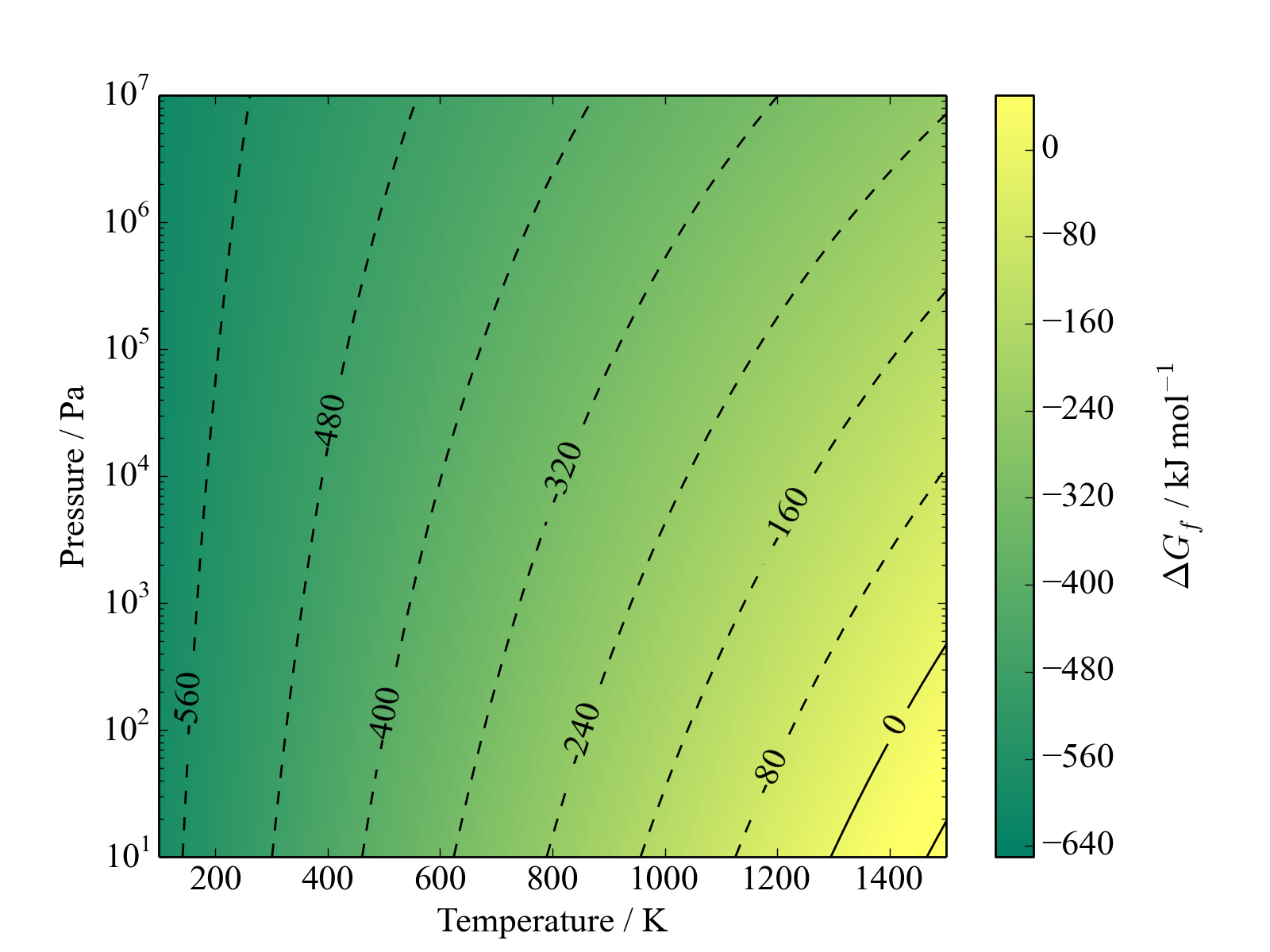} \\
(a) $\ce{2Cu} + \ce{Zn} + \ce{Sn} + \ce{2S2} \rightleftharpoons  \ce{Cu2ZnSnS4}$
\includegraphics[width=\columnwidth]{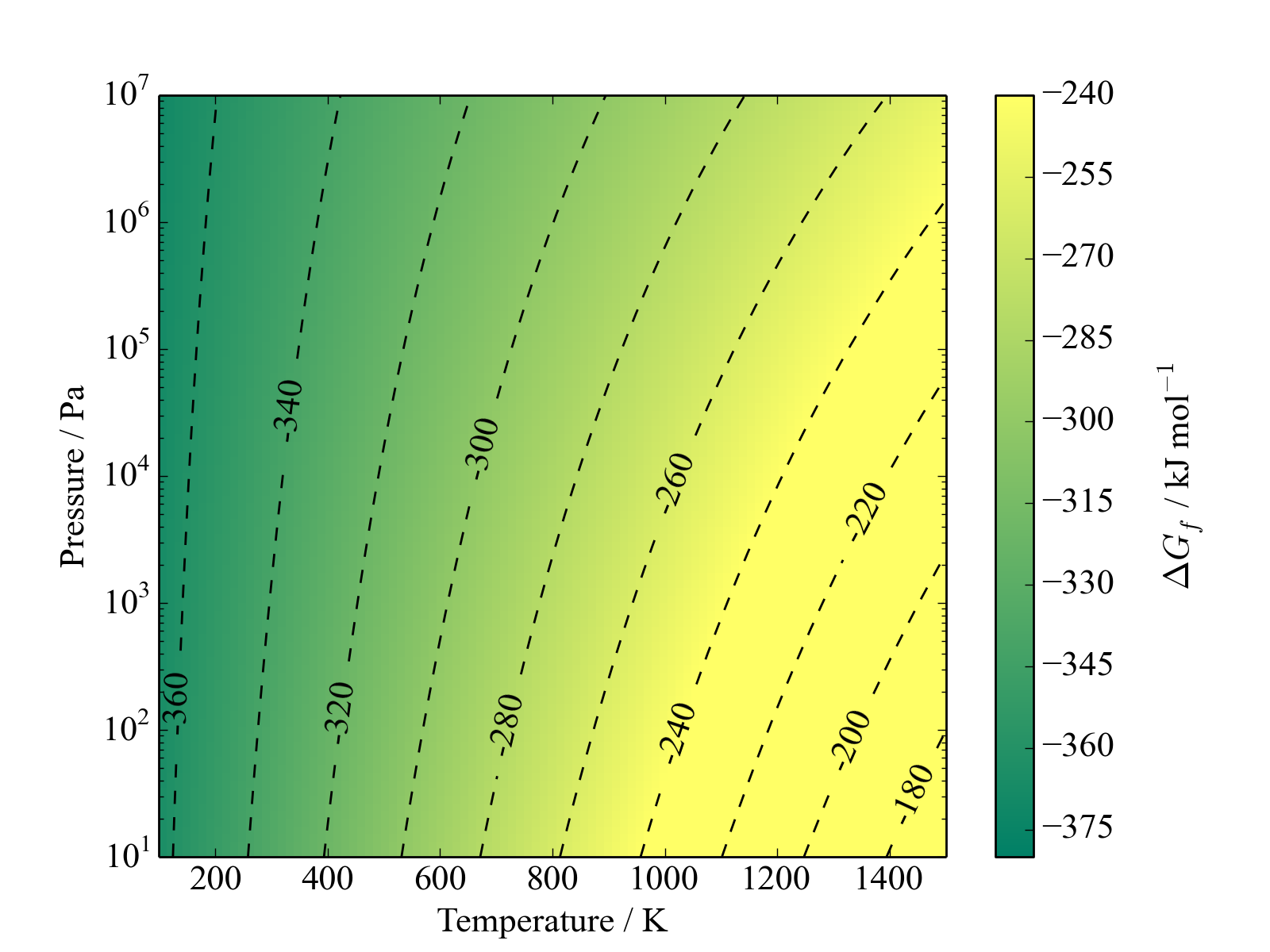}
(b) $\ce{2Cu} + \ce{Zn} + \ce{Sn} + \tfrac{1}{2}\ce{S8} \rightleftharpoons \ce{Cu2ZnSnS4}$
\end{center}
\caption{Gibbs free energy of formation of kesterite CZTS from metals and sulfur gases.
Contours represent energies in kJ~mol$^{-1}$.
The y-axis represents a partial pressure of gaseous sulfur in the form of (a) \ce{S2} and (b) \ce{S8}.
\label{fig:DG_CZTS_Sx}
}
\end{figure}
It is clear that CZTS is thermodynamically stable with respect to its solid elemental precursors over all reasonable processing conditions.
Considering the equilibrium with sulfur vapours:
\begin{align}
\ce{2Cu} + \ce{Zn} + \ce{Sn} + \ce{2S2}\text{(g)} &\rightleftharpoons \ce{Cu2ZnSnS4} \\
\ce{2Cu} + \ce{Zn} + \ce{Sn} + \tfrac{1}{2}\ce{S8}\text{(g)} &\rightleftharpoons \ce{Cu2ZnSnS4}
\end{align}
we see a stronger interaction at high temperatures and low pressures (Fig.~\ref{fig:DG_CZTS_Sx}).
This is driven by entropy, as it is more entropically favourable for sulfur to enter a low-pressure environment.
The effect is greatest for \ce{S2}, suggesting an instability at high temperatures over 1300~K at low pressures,
whereas if only \ce{S8} is to be considered then the formation appears irreversible even under relatively extreme conditions.
Given that the actual composition of sulfur vapours is known to shift towards \ce{S2} at high temperatures, Fig.~\ref{fig:DG_CZTS_Sx}a is more appropriate in this regime.\cite{Rau1973a}

\subsection{Stability of CZTS relative to binary sulfides}

\begin{figure}
\includegraphics[width=\columnwidth]{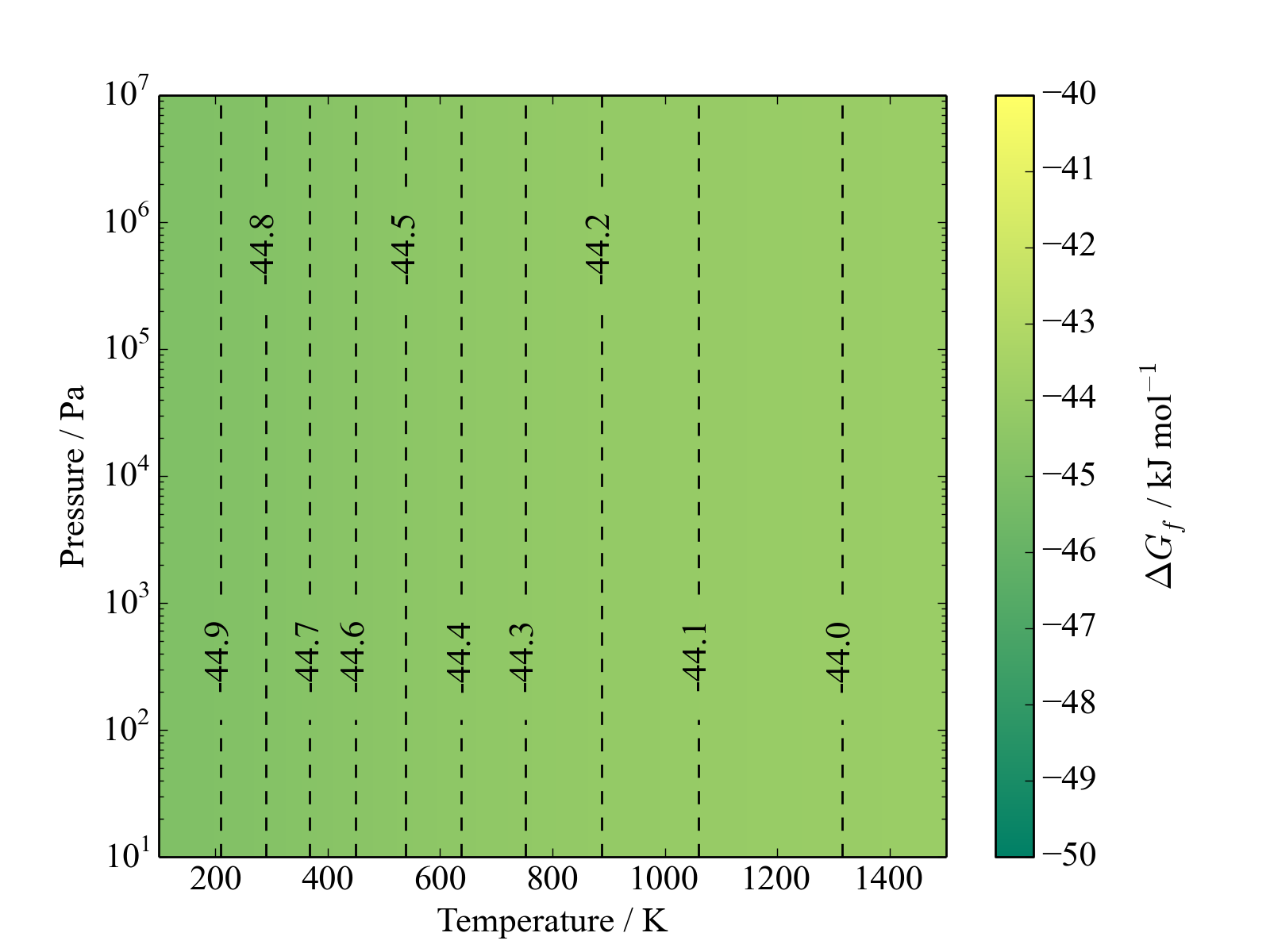}
\caption{Gibbs free energy of formation of kesterite from binaries (Eqn. \ref{eqn:binary}).
Contours represent energies in kJ~mol$^{-1}$.
The y-axis represents pressure from an inert gas or mechanical force.
\label{fig:DG_CZTS_binaries}
}
\end{figure}

The binary sulfides are of interest both in terms of routes to forming CZTS and possible disproportionation reactions. 
Note that for a stoichiometric mixture, there is no dependence on the chemical potential of elemental sulfur:
\begin{equation}\label{eqn:binary}
\ce{Cu2S} + \ce{ZnS} + \ce{SnS2} \rightleftharpoons \ce{Cu2ZnSnS4},
\end{equation}
and hence the temperature-pressure dependence is again very mild (Fig.~\ref{fig:DG_CZTS_binaries}).
However, the overall free energy change is considerably reduced to the order $-44$~kJ~mol$^{-1}$; this is logical as the binary phases are themselves stable with respect to their elemental precursors.

\begin{figure}
\includegraphics[width=\columnwidth]{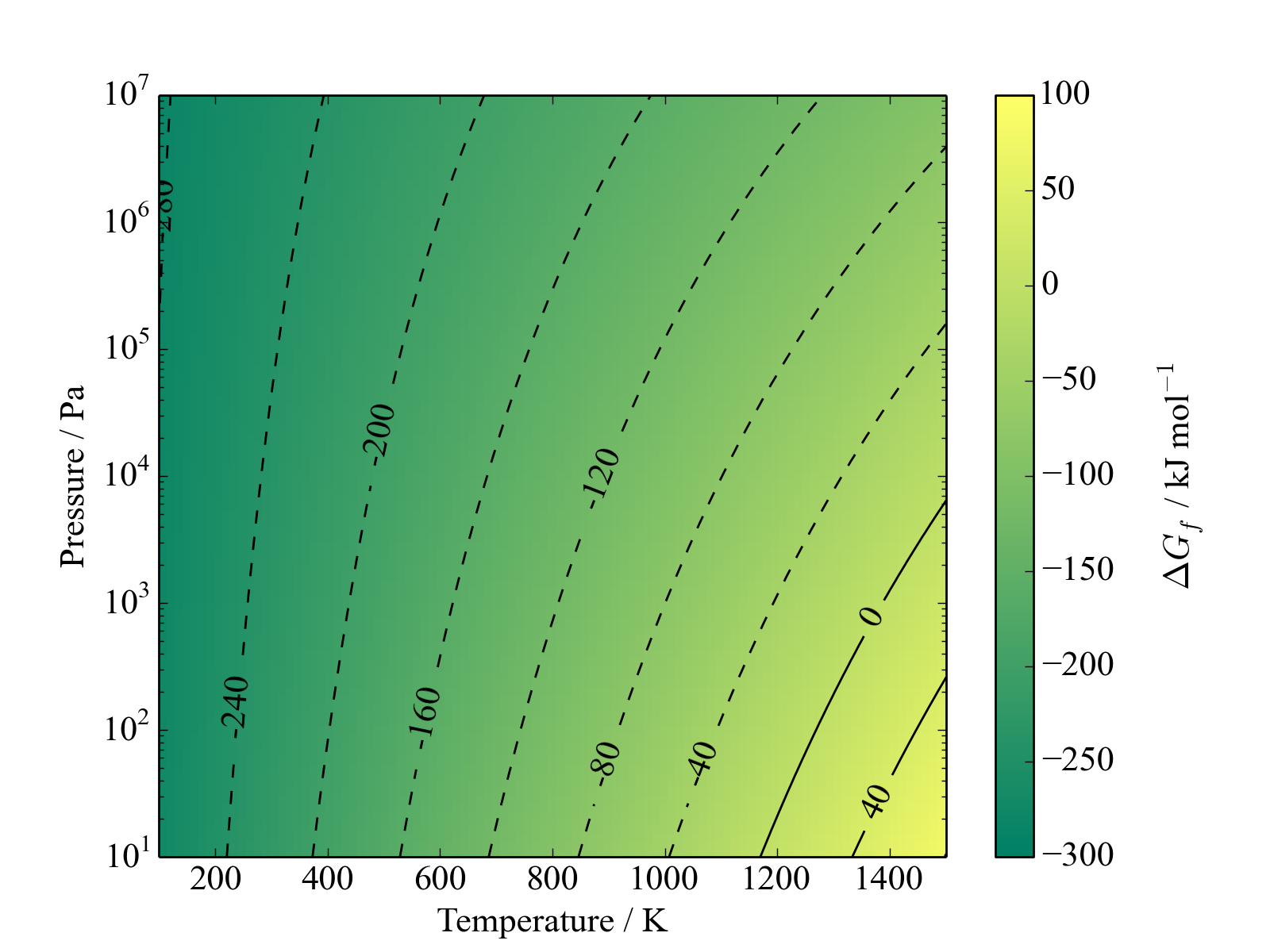}
\caption{Gibbs free energy of formation of kesterite from binaries, tin and sulfur vapour (Eqn. \ref{eqn:binary-sn}).
Contours represent energies in kJ~mol$^{-1}$.
The y-axis represents the partial pressure of \ce{S2}.
\label{fig:DG_CZTS_SnS2}
} 
\end{figure}

If we instead consider decomposition to Sn metal, due to an instability of \ce{SnS2}, then a dependence appears as sulfur is released (Fig.~\ref{fig:DG_CZTS_SnS2}) following the reaction
\begin{equation}\label{eqn:binary-sn}
\ce{Cu2ZnSnS4} \rightleftharpoons \ce{Cu2S} + \ce{ZnS} + \ce{Sn} + \ce{S2}\text{(g)}.
\end{equation} 
This relationship is even stronger in the event of partial sulfur loss to form the divalent tin monosulfide, SnS:
\begin{equation}\label{eqn:binary-sns}
\ce{Cu2ZnSnS4} \rightleftharpoons \ce{Cu2S} + \ce{ZnS} + \ce{SnS}\text{(s)} + \tfrac{1}{2}\ce{S2}\text{(g)}.
\end{equation}
Note that Eqn.~\ref{eqn:binary-sns} appears identical to Eqn.~\ref{eqn:scragg-SnS};
the only difference is that SnS is here defined as being the bulk solid,
while it is understood that in the actual mechanism the SnS likely forms a reactive surface and may even be adsorbed to a CZTS or ZnS bulk phase.
\begin{figure}[hb!]
\includegraphics[width=\columnwidth]{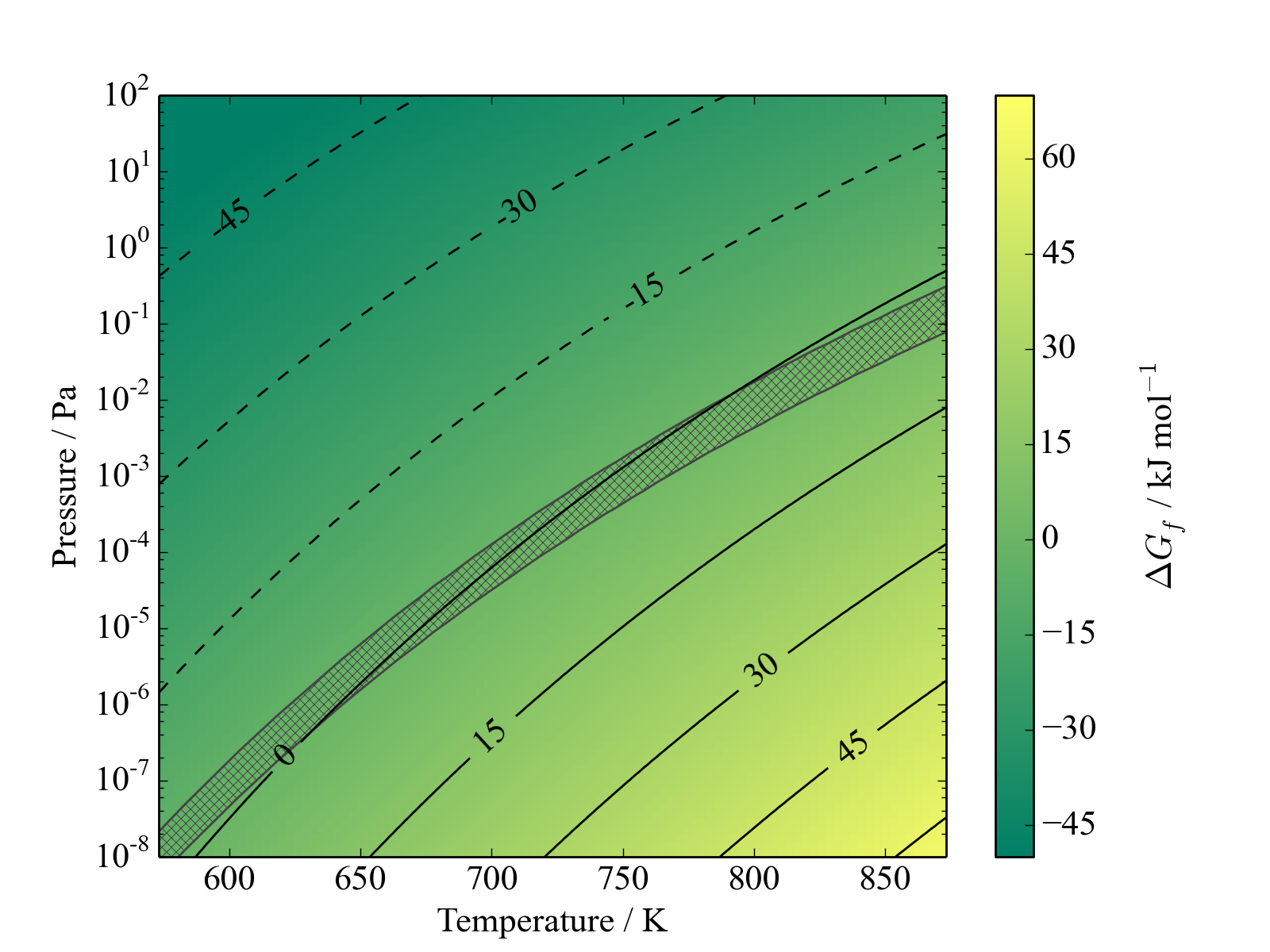} 
\caption{Gibbs free energy of formation of kesterite from binary compounds, with SnS in equilibrium with \ce{S2} vapours (Eqn. \ref{eqn:binary-sns}).
Contours represent energies in kJ~mol$^{-1}$.
The y-axis represents the partial pressure of \ce{S2}.
Hatched area shows predicted transition region from kinetic modelling, figure 5 of Ref.~\citenum{Scragg2011a}.
Note low pressure and reduced temperature range relative to other Figures; this is for ease of comparison with Ref.~\citenum{Scragg2011a}.
\label{fig:DG_CZTS_SnS}
}
\end{figure}
Application of the model at high ($\sim$1~bar) partial pressures of sulfur suggest that such decomposition is not expected below temperatures of around 1300K.
Figure~\ref{fig:DG_CZTS_SnS} shows that CZTS becomes unstable following this process at around 600K and above for very low partial pressures of \ce{S2}. 
The stability window is found to be almost identical to that predicted by kinetic modelling of experimental data (see figure 5 of Ref.~\citenum{Scragg2011a}).

The agreement is especially interesting given that the kinetic model of \citeauthor{Scragg2011a} is based on SnS vapours (which were introduced as a gas stream in the accompanying experiment),
while the result is reproduced here in an \emph{ab initio} thermodynamic model with no such phase.\cite{Scragg2011a}
This model does not consider the evaporation of solid SnS, and as such is equivalent to the two-step model (Eqns.~\ref{eqn:scragg-SnS}-\ref{eqn:SnS-vapour}) given a saturated SnS vapour phase.
It is reasonable to expect that the direct formation of SnS vapour would offer a higher entropy gain, lowering the free energy further and hence promoting decomposition; this would be equivalent to same two-step model with a very low partial pressures of \ce{SnS}.

Direct comparison to experimental syntheses of CZTS is difficult as stability curves have not been measured directly, but it is possible to interpret experimental results where the conditions are reported clearly.
At typical formation temperatures of 700-800~K we find that decomposition is only expected where there is a significant absence of sulfur in the atmosphere.
Given that sulfur solids have a vapour pressure of the order 100 kPa in this temperature range, this would only be expected to occur where sulfur is limited or vapours are removed by a vacuum pump.\cite{crc}
Ericson \emph{et al.} successfully produced CZTS films by reactive sputtering followed by annealing in a static argon atmosphere of 35~kPa at 560$^\circ$; 
given that they observed a correlation between sulfur loss and temperature during sputtering, we would assume that the sulfur was sufficiently mobile to form an equilibrium pressure during annealing.\cite{Ericson2013}
Redinger \emph{et al.} observed decomposition at 560~$^\circ$C under ``vacuum''; this is in agreement with our model provided that their vacuum pump maintained a sulfur partial pressure of around 0.1 Pa ($10^{-3}$~mbar) or less.\cite{Redinger2011a}

\section{Conclusions}
Based on first-principles total energy calculations, a thermodynamic model has been developed to describe the formation and stability of CZTS with respect to its elemental constituents and stoichiometric binary sulfides, as well as the tin monosulfide which is known to play an active role. 
Reactions involving solid and gaseous sulfur have been considered, the latter of which introduces a substantial temperature and pressure dependence. 
Temperature and pressure conditions have been related to the phase equilibrium; they indicate higher decomposition temperatures than those observed experimentally, but otherwise broadly similar behaviour.
These results, which are \emph{ab initio} except for the reference data for sulfur vapours, closely reproduce a previous model which was derived from experimental results and reference data.\cite{Scragg2011a}
It is clear that the pressure-temperature interaction is strong in near-vacuum conditions.

This initial model is based on a number of approximations that could be removed in future work. 
These include the effects of thermal expansion and compressibility on the solid phases and the non-ideality of the S vapour. 
There is always a compromise between accuracy and computational cost, especially as future models will be extended to consider competitive ternary and quaternary phases, in addition to metallic alloys.
One issue with rigorous validation of the model is the scarcity of experimental thermodynamic data so far. 
%
We suggest that high-temperature near-vacuum experiments need to control and report the annealing pressure carefully if they are to be reproducible and aid understanding of the phase equilibria of multi-component semiconductors such as \ce{Cu2ZnSnS4}.


  
\section{Acknowledgements}
We acknowledge L.~M.~Peter for useful discussions and J.~J.~Scragg for providing access to his kinetic modelling data. 
A.W. acknowledges support from the Royal Society for a University Research Fellowship and A.J.J. is funded by EPSRC (Grant No. EP/G03768X/1).
We acknowledge use of Hartree Centre resources in this work. The STFC Hartree Centre is a research collaboratory in association with IBM providing High Performance Computing platforms funded by the UK's investment in e-Infrastructure. 
Via our membership of the UK's HPC Materials Chemistry Consortium, which is funded by EPSRC (EP/L000202), this work made use of the facilities of HECToR, the UK's national high-performance computing service, which is provided by UoE HPCx Ltd at the University of Edinburgh, Cray Inc and NAG Ltd, and funded by the Office of Science and Technology through EPSRC's High End Computing Programme.




\footnotesize{
\bibliography{thebib} 
\bibliographystyle{rsc} 
}

\end{document}